\documentclass[aps, amsfonts, amsmath, nofootinbib, showpacs, byrevtex, preprintnumbers, showkeys, preprint, prd]{revtex4}   
\usepackage{epsf}
\usepackage{graphicx}
\usepackage{bbm}
\newcommand{\bi}{\bigskip}
\newcommand{\no}{\noindent}
\newcommand{\be}{\begin{eqnarray}}
\newcommand{\ee}{\end{eqnarray}}
\newcommand{\hk}{\hspace{0.1cm}}

\newcommand{\rk}{\right)}
\newcommand{\lk}{\left(}

\newcommand{\sli}{\sum\limits}

\newcommand{\il}{\int\limits}

\newcommand{\vx}{\vec{x}}
\newcommand{\vy}{\vec{y}}
\newcommand{\vz}{\vec{z}}
\newcommand{\mybf}{}

\newcommand{\ingrscale}{1.8}

\begin{document}

\title{The 't~Hooft loop in the Hamiltonian approach to Yang-Mills theory in Coulomb gauge}\thanks{Supported in part by DFG under
Re 856/6-1 and Re 856/6-2 and by the European Graduate School
Basel-Graz-T\"ubingen.}
\author{H. Reinhardt and D. Epple}
\affiliation{Institut f\"ur Theoretische Physik\\
Auf der Morgenstelle 14\\
D-72076 T\"ubingen\\
Germany}


\date{\today}
\begin{abstract}
The spatial 't~Hooft loop, which is a disorder parameter dual to the temporal
Wilson loop, is calculated  using the nonperturbative Yang-Mills
vacuum wave functional determined previously by a variational solution of the
Yang-Mills Schr\"odinger equation in Coulomb gauge. 
It is shown, that this wave functional yields indeed a
perimeter law for large spatial 't~Hooft loops signaling confinement.
\end{abstract}
\maketitle
\bi



\no
\section{Introduction}
\bi

\no
Order and disorder parameters are useful tools in determining the phase
structure of extended physical systems. The order parameter of confinement in
QCD is the temporal Wilson loop, which shows an area law in the confinement
regime and a perimeter law in the deconfinement region \cite{Wilson:1974sk}. 
The temporal Wilson loop is
related to the potential between static quarks and an area law corresponds to a
linearly rising quark potential.
\bi

\no
The emergence of the area and perimeter laws in the Wilson loop in the confined
and deconfined
 phases can be easily understood in a center vortex picture of the
Yang-Mills vacuum \cite{Mack:1980rc}. 
While confinement arises for a phase of percolated vortices,
the perimeter law arises from small vortex loops, which can link to the Wilson
loop only if they are in the vicinity of the loop. In recent years the center vortex picture of confinement has
received strong support from lattice calculations \cite{DelDebbio:1996mh,Engelhardt:1999fd}.
\bi

\no
Another (dis-)order parameter for Yang-Mills theory was introduced by 't~Hooft
\cite{'tHooft:1977hy}
and is defined in the following way. Consider Yang-Mills theory in canonical
quantization (which assumes Weyl gauge 
$A_0 = 0$), where the spatial components of the
gauge field $\vec{A} (x)$ are the ``coordinates''. Let $W [A] (C)$ denote the
spatial Wilson loop operator. Then the operator of 't~Hooft's disorder parameter $V (C)$
for a closed (spatial) loop $C$ is defined by the commutation relation 
\be
V (C_1) W (C_2) = Z^{L (C_1, C_2)} W (C_2) V (C_1)  \hk ,
\label{1}
\ee
where $Z$ is a (non-trivial) center element of the gauge group and $L (C_1,
C_2)$ is the Gaussian linking number of the two loops $C_1$ and $C_2$. The
expectation value of the (spatial) 't~Hooft loop operator $V (C)$ can serve as
an order parameter for Yang-Mills theory. As argued by 't~Hooft, in the
confined (deconfined) phase $\langle V (C) \rangle$ obeys a perimeter (area)
law \cite{'tHooft:1979uj}. In this sense $\langle V (C) \rangle$ behaves dual to the temporal Wilson
loop, which shows an area (perimeter) law in the confined (deconfined) phase.
Given the fact, that a center vortex located at a loop $C_1$ and described by a
gauge potential ${\cal A} (C_1)$ produces  a Wilson loop
$C_2$
\be
\label{2}
W [{\cal A} (C_1)] ( C_2) = Z^{L (C_1, C_2)} \hk ,
\ee
 the 't~Hooft loop operator $V (C)$ defined by equation
(\ref{1}) can be interpreted as a center vortex creation operator. An
analogous monopole creation operator was studied on the lattice in 
ref.\ \cite{DelDebbio:1995yf}. 
\bi

\no
In statistical physics, operators creating topological excitations like vortices
or monopoles are referred to as disorder operators. Their expectation values,
referred to as disorder parameters,
 are related to the free energy of the associated
topological excitations. 
\bi

\no
Center vortices can be easily generated on a lattice, where they represent
co-closed $(D - 2)$-dimensional hyper-surfaces of plaquettes equal to a non-trivial
center element. Accordingly, the 't~Hooft loop operator can be easily realized
on the lattice. In particular, center vortices wrapping around the
whole lattice universe (torus), can be easily generated, and 't~Hooft loops of
maximal size on a finite lattice can be realized by imposing twisted boundary
condition on the links \cite{'tHooft:1979uj}.
 The free energy of center vortices wrapping around the 
(space-time) 
torus, i.e.\ 't~Hooft loops of maximal size,
 has been calculated in a high-temperature expansion \cite{Munster:1980iv} and measured on the lattice for $SU (2)$ Yang-Mills theory in ref.\
\cite{Kovacs:2000sy} 
and for a $Z (3)$ random center vortex model \cite{Engelhardt:2003wm} 
in ref.\ \cite{Quandt:2004gy}. Let us also mention that the energy per length of
a straight infinite magnetic vortex line was calculated to one loop order as
a function of the vortex flux \cite{Lange:2003ti}. It was found that the center vortex flux
is energetically favored and its (free) energy vanishes in the absence of
quarks. 
\bi

\no
In the present paper we calculate the (spatial) 't~Hooft loop, i.e.\ the vacuum
expectation value of the operator $V(C)$ defined by eq.\ (\ref{1}), in continuum Yang-Mills theory. 
Recently progress has been made in variationally solving the
Yang-Mills Schr\"odinger equation in Coulomb gauge \cite{Szczepaniak:2001rg,
Feuchter:2004mk,Reinhardt:2004mm,Epple:2006hv}. The
Coulomb gauge has the advantage that Gauss' law can be explicitly resolved
resulting in an explicit expression for the potential between static (color)
charges. Using Gaussian ans\"atze for the vacuum wave functional,
minimizing the vacuum energy density yields a set of Dyson-Schwinger
equations (DSEs). These equations can be solved analytically in the infrared 
\cite{Schleifenbaum:2006bq}.
Implementing the so-called horizon condition (i.e.\ an infrared diverging ghost
form factor) one finds a solution which produces a strictly linearly rising
confinement potential for static color charges, provided the curvature of the
space of gauge orbits, induced by the Faddeev-Popov determinant, is properly
included. In this paper we use the Yang-Mills vacuum wave functional determined
in refs.\ \cite{Feuchter:2004mk,Epple:2006hv}
 to calculate the spatial 't~Hooft loop.
\bi

\no
The organization of the paper is as follows: In the next section we present the
't~Hooft loop in Coulomb gauge, and the loop geometry will be worked out for a
planar 
circular loop.
We will briefly review the Dyson-Schwinger equations in Coulomb gauge derived in ref.\ \cite{Feuchter:2004mk} in sect.\ \ref{sec-revisited}.
In sect.\ \ref{sec-asymptotic} we study the asymptotic behavior for the 't~Hooft
loop in both the ultraviolet and infrared and extract its large perimeter 
behavior.
Our numerical results will be presented in sect.\ \ref{sec-results}. A short summary and some
concluding remarks are given in sect.\ \ref{sec-concl}.
\bi

\no
\section{The spatial 't~Hooft loop in Coulomb gauge}\label{sec-loop}

\subsection{The 't~Hooft loop operator}

\no
The 't~Hooft loop operator, $V(C)$, is implicitly defined by eq.\ (\ref{1}).
An explicit realization of 't~Hooft's
loop operator $V(C)$ in continuum Yang-Mills theory was derived in ref.\ 
\cite{Reinhardt:2002mb} and
is given by
\be
\label{3}
V (C) = \exp \left[ i \int d^3 x {\cal A}^a_i (x) \Pi^a_i (x) \right] \hk ,
\ee
where $\Pi^a_i (x) = - i \delta / \delta A^a_i (x)$ is the canonical momentum
operator and
\be
\label{4}
{\cal A}^a_i (\Sigma; x) = \xi^a \il_\Sigma d^2 \tilde{\sigma}_i 
\delta^3 (x - \bar{x}
(\sigma)) \hk 
\ee
is the gauge potential of a center vortex, whose flux is localized at the loop
$C = \partial \Sigma$. Here $\xi = \xi^a T_a$ (with $T_a$ being the generators
of the gauge group) denotes a co-weight vector defined by $\exp (- \xi) = Z$
with $Z$ being a (non-trivial) center element of the gauge group. 
Furthermore $\bar{x}_i (\sigma)$ denotes a parameterization of the
2-dimensional surface $\Sigma$ bounded by the loop $C$.
For $SU (2)$ with generators $T_a = - \frac{i}{2} \tau_a$ ($\tau_a$ being the
Pauli matrices) the co-weight vector of the non-trivial center element $Z = - 1$
is given by $\xi^a = \delta^{a 3} 2 \pi$.

Just as the Wilson loop $W (C_2)$ creates an elementary
electric flux along the loop $C_2$, the 't~Hooft loop $V (C_1)$ creates an
elementary magnetic flux along $C_1$. In fact, the gauge potential (\ref{4}) produces a magnetic flux
\be
\label{5}
\vec{B} (x) = \vec{\partial} \times
 \vec{\cal A} (\Sigma, x) = \xi \oint\limits_{\partial
\Sigma} d \vec{y} \delta^3 (\vec{x} - \vec{y}) \hk ,
\ee
which is localized on $C=\partial \Sigma$.
\bi

\no
Inserting (\ref{4}) into (\ref{3}), the spatial integral can be carried out yielding 
\begin{align}
V(C)=\exp\left[i\xi^a\int\limits_\Sigma\,d^2\tilde{\sigma}_i\Pi^a_i(\bar{x}(\sigma))\right]\, .
\end{align}
Since $\Pi^a_i(x)$ is the operator of the electric field, it is seen that the 't~Hooft loop measures the electric flux through the surface $\Sigma$ enclosed by the loop $C=\partial\Sigma$, in the same way as the (spatial) Wilson loop $W(C)$ measures the magnetic flux through a surface enclosed by the loop $C$.
\bi

\no
The loop operator (\ref{3}) is not manifestly gauge invariant. However, it
produces a gauge invariant result, when it acts on a gauge invariant wave
functional, see ref.\ \cite{Reinhardt:2002mb} for more details. 
\bi

\no
Consider the action of the loop operator (\ref{3}) on a gauge
invariant wave functional $\Psi [A]$. The 't~Hooft loop operator is analogous to
the translation operator $e^{i a \hat{p}}$ in quantum mechanics which shifts the
argument of a wave function $\exp (i a \hat{p}) \varphi (x) = \varphi (x + a)$. 
Thus, when acting on wave functionals the loop operator (\ref{3}) shifts the
``coordinate'' $\vec{A} (x)$ by the center vortex field ${\cal A} (x)$
\be
\label{7}
V (C = \partial \Sigma) \Psi [A] = \Psi [ A + {\cal A} (\Sigma)] \hk 
\ee
and in this sense is a true center vortex generator. 
\bi

\no
Below, we firstly derive the expression for the 't~Hooft loop in Coulomb 
gauge and
secondly calculate the 't~Hooft loop for the specific vacuum wave functional
obtained in refs.\ \cite{Feuchter:2004mk,Reinhardt:2004mm,Epple:2006hv}.
Furthermore we will explicitly work out the 't~Hooft loop
for a planar circular loop.
\bi

\no
\subsection{The vacuum expectation value of the 't~Hooft loop operator}
\bi

\no
We are interested in the vacuum expectation value of the 't~Hooft loop operator (\ref{3}).
The exact vacuum wave functional is not known but recently
progress has been made in determining the Yang-Mills vacuum wave functional
$\Psi_0 (A)$ variationally in Coulomb gauge $\vec{\partial} \vec{A} = 0$
\cite{Szczepaniak:2001rg, 
Feuchter:2004mk,Reinhardt:2004mm,Epple:2006hv}. The
wave functionals in Coulomb gauge satisfy Gauss' law\footnote{Gauss' law is 
explicitly
resolved in deriving the Yang-Mills Hamiltonian in Coulomb gauge.} and hence
should be regarded as the gauge invariant wave functionals restricted to
transverse gauge fields. 
Using
eq.\ (\ref{7}) 
and implementing the Coulomb gauge by means of the Faddeev-Popov
method, we obtain for the expectation value of the 't~Hooft loop operator in
Coulomb gauge wave functionals
\be
\label{8}
\langle V (\partial \Sigma) \rangle & = & \langle \Psi | V (\partial \Sigma) |
\Psi \rangle \nonumber\\
& = & \int D A^\perp J (A^\perp) \Psi^* (A^\perp) \Psi (A^\perp + {\cal A}^\perp
(\Sigma)) \hk ,
\ee
where 
\be
\label{9}
J (A^\perp) = Det (- \hat{D} \partial) 
\ee
is the Faddeev-Popov determinant with
$\hat{D}$ being the covariant derivative in the adjoint representation.\footnote{The Fadeev-Popov determinant provides the Haar measure of the gauge group. \cite{Reinhardt:1996fs}}
\bi

\no
Due to the transversal property of the field variable $A^\perp_i (x)$ 
only the transversal part ${\cal A}^\perp (\partial \Sigma ; x)$ 
of the center vortex gauge potential (\ref{4}) enters, which is
given by \cite{Reinhardt:2001kf}
\be
\label{16}
{\cal A}^\perp_i (\partial \Sigma , x) = - \xi \oint\limits_{\partial \Sigma} d
\tilde{\sigma}_{i k} \partial^{\bar{x}}_k D (x - \bar{x} (\sigma)) \hk , \hk d
\tilde{\sigma}_{i k} = \epsilon_{i k l} d \bar{x}_l
\ee
where $D (x)$ is the Green's function of the 3-dimensional Laplacian defined by
\be
\label{17}
- \Delta_x D (x) = \delta^3 (x) \hk .
\ee
Note, that the transversal part ${\cal A}^\perp (\partial
\Sigma ; x)$ depends manifestly only on the boundary $\partial
\Sigma$\footnote{The longitudinal part of ${\cal A} (\Sigma, x)$ is given by
\cite{Reinhardt:2001kf}
\be
\label{18}
{\cal A}^{||} (\Sigma, x) = - \xi \partial \Omega (\Sigma, x) \hk ,
\ee
where
\be
\label{19}
\Omega (\Sigma, x) = \il_\Sigma d^2 \tilde{\sigma}_k \partial^x_k D (x - \bar{x}
(\sigma)) 
\ee
is the solid angle  subtended by the surface $\Sigma$ at the point $x$.}.
\bi

\no
In ref.\ \cite{Feuchter:2004mk} 
the following ansatz was used for the vacuum wave functional 
\be
\label{10}
\Psi [A^\perp] & = & 
{\cal N} \frac{1}{\sqrt{J (A^\perp)}} \exp \lk - S_0 [A^\perp] \rk
\hk , \hk \nonumber\\
 S_0 [A^\perp] & = & \frac{1}{2}
\int d^3 x d^3 x' A^{\perp a}_i (x) t_{i j} (x) \omega (x, x') A^{\perp a}_j
(x')  \hk ,
\ee
where $t_{i j} (x) = \delta_{i j} - \partial^x_i \partial^x_j/\partial^2_x$
denotes the transversal projector and 
$\omega (\vec{x}, \vec{x'})$ is a variational kernel determined
by minimizing the vacuum energy density. The ansatz (\ref{10}) for the 
wave functional is motivated by the form of the wave function
of a point particle in a spherically symmetric $(l =
0)$ s-state, which is of the form $\Psi (r) = \phi (r) / r$, where $J = r^2$
is the Jacobian corresponding to the change from Cartesian coordinates to
spherically symmetric coordinates (For s-states the scalar product is given by
$\langle \Psi_1 | \Psi_2 \rangle = \int d r r^2 \Psi^*_1 (r) \Psi_2 (r)$). Like
in the case of the point particle in a spherically symmetric state, the ansatz
(\ref{10}) simplifies the actual calculation, since it removes the Jacobian from
the integration measure. Let us stress, however, that both, the ultraviolet and
the infrared properties of the theory do not depend on the specific choice of
the pre-exponential factor. In fact 
it was shown \cite{Reinhardt:2004mm}, that
replacing $J^{- \frac{1}{2}}$ by $J^{- \alpha}$ with an arbitrary real $\alpha$
yields, after minimizing the energy density (to 2 loop level), a vacuum wave
functional whose infrared limit is independent of the choice of $\alpha$. (The
choice $\alpha = \frac{1}{2}$ simplifies, however, the calculations.) 
\bi

\no
For the
wave functional (\ref{10}) the expectation value of the 't~Hooft loop (\ref{8})
becomes 
\be
\label{11}
\langle V (\partial \Sigma) \rangle & = & | {\cal N} |^2 \int {\cal D} A^\perp J^{\frac{1}{2}} [A^\perp] J^{-
\frac{1}{2}} [A^\perp + {\cal A}^\perp] \exp \lk - \lk S_0 [A^\perp] + S_0 
[A^\perp + {\cal
A}^\perp] \rk \rk \hk .
\ee
In ref.\ \cite{Reinhardt:2004mm} it was shown, that to 2-loop level in the energy the
Jacobian can be expressed as
\be
\label{12}
J (A) = \mathrm{const}\cdot\exp \lk - \int d^3 x d^3 x' A^{\perp a}_i (x) t_{i j} (x) \chi (x,
x') A^{\perp a}_j (x') \rk\hk ,
\ee
where \cite{Feuchter:2004mk}
\be
\label{13}
\chi (x, y) = -\frac{1}{4} \frac{1}{N^2_C - 1} \delta^{a b} t_{k l} (x)   
\left\langle \Psi \left| \frac{\delta^2 \ln J
(A^\perp)}{\delta A^{\perp a}_k (x) \delta A^{\perp b}_l (y)} \right| \Psi
\right\rangle 
\ee
is the ``curvature'' of the space of gauge orbits. This quantity gives the
ghost loop contribution to the gluon self-energy. The ghost propagator is defined by
\begin{align}
\label{ghost-propagator}
\langle\psi|\langle\vec{x}|(-\hat{D}\partial)^{-1}|\vec{x}'\rangle|\psi\rangle = \langle\psi|\frac{d}{-\Delta}|\psi\rangle
\end{align}
where $d$ denotes the ghost form factor.

Using the representation
(\ref{12}) the functional integral in eq.\ (\ref{11}) becomes Gaussian and we
obtain
\be
\label{14}
\langle V (\partial \Sigma) \rangle = \exp \left[ - \frac{1}{2} \int d^3 x d^3 y
{\cal A}^{a \perp}_i (\Sigma ; x) t_{i j}(x) K 
(x, y) {\cal A}^{\perp a}_j (\Sigma ; 
y) \right] \equiv \exp
(- S) \hk ,
\ee
where we have introduced the abbreviation
\be
\label{15}
K (x, y) = \omega (x, y) - \chi (x, y) - \frac{1}{2} \int d^3 z d^3 z' (\omega
(x, z) - \chi (x, z)) \omega^{- 1} (z, z') \lk \omega (z', y) - \chi (z', y) \rk
\hk .
\ee
If one uses the more general ansatz for the vacuum wave functional considered in
ref.\ \cite{Reinhardt:2004mm},
\be
\label{X1}
\Psi (A) = J (A)^{- \alpha} e^{- \frac{1}{2} \int A \omega A},
\ee
and uses the representation (\ref{12}) for the Jacobian, one still finds the form (\ref{14}) for the 't
Hooft loop, however, with the kernel
\be
\label{X2}
K = \Omega - \chi - \frac{1}{2} (\Omega - \chi) \Omega^{- 1} (\Omega - \chi) \hk
,
\ee
where
\be
\label{X3}
\Omega = \omega + (1 - 2 \alpha) \chi \hk 
\ee
is the inverse (3-dimensional) gluon propagator.
Note, eq.\ (\ref{X2}) arises from eq.\ (\ref{15}) with $\omega$ replaced by
$\Omega$ (\ref{X3}). In ref.\ \cite{Reinhardt:2004mm} it was shown, that to one loop level
(i.e.\ 2 loops in the energy density) the solutions of the Schwinger-Dyson
equations for $\Omega$ and $\chi$ are independent of the choice of $\alpha$.
Thus, we obtain the same expectation value for the 't~Hooft loop for all choices
of $\alpha$ and for convenience we will continue to choose $\alpha =
\frac{1}{2}$ for which $\Omega = \omega$. We should, however, stress that
 there are
other quantities (like e.g.\ the 3-gluon vertex), which are sensitive to the
choice of $\alpha$ \cite{Schleifenbaum:2006bq}.
\bi

\no
Using eq.\ (\ref{16}) one finds after straightforward evaluation for the exponent
in the 't~Hooft loop (\ref{14})
\be
\label{20}
S 
& = & 2 \pi^2 \oint\limits_C y_k \oint\limits_C d z_l \int d^3 x
t_{k l} (\vec{y}) K 
(\vec{y} , \vec{x}) D (\vec{x} , \vec{z}) \hk .
\ee
Due to the translational invariance of the vacuum the kernel $K (\vx, \vy)$, as well as
the Green function $D (\vx, \vy)$, depends only on the distance $| \vx - \vy|$
and it is convenient to use Fourier representation
\be
\label{21}
D (\vec{x} - \vec{x'}) & = & 
\int \frac{d^3 q}{(2 \pi)^3} e^{i \vec{q} (\vec{x} - \vec{x'})} D (q) \hk , \hk D
(q) = \frac{1}{q^2} \hk , \hk  t_{k l} (\hat{q}) = \delta_{kl} -
\hat{q}_k \hat{q}_l \hk , \hk \hat{q} = \frac{\vec{q}}{| \vec{q} |} \hk .
\ee
Then the kernel (\ref{15}) becomes
\be
\label{6-24}
K (q)  & = & \lk \omega (q) - \chi (q) \rk 
\left[ 1 - \frac{1}{2} \frac{\omega (q) - \chi
(q)}{\omega (q)} \right]  = \frac{1}{2} \omega (q) \lk 1 - \lk \frac{\chi
(q)}{\omega (q)} \rk^2 \rk  \hk .
\ee
We obtain then from eq.\ (\ref{20}) 
\be
\label{22}
S (C) = \il^\infty_0 d q K (q) H (C, q) \hk ,
\ee
where
the function
\be
\label{23}
H (C; q) = \frac{1}{4 \pi} \oint\limits_C d y_k \oint\limits_C d z_l \int d
\hat{\Omega}_q t_{k l} (\hat{q}) e^{i \vec{q} \cdot (\vec{y} - \vec{z})}
\ee
contains the geometry of the loop $C$ considered, but is independent of the
properties of the Yang-Mills vacuum, which are exclusively
 contained in the kernel
$K$. 
\bi

\no
\subsection{Loop Geometry}
\bi

\no
We can explicitly carry out the integral over the solid angle $\int d
\hat{\Omega}_q$  in (\ref{23}). 
Expressing the unit vector $\hat{q}$ in terms of spherical
coordinates $\Theta, \phi$
\be
\label{24}
\hat{q} = \hat{q} (\Theta, \phi) = \lk \sin \Theta \cos \phi , \sin \Theta \sin
\phi, \cos \Theta \rk \hk 
\ee
we have
\be
\label{25}
\il^{2 \pi}_{0} d \phi t_{k l} (\hat{q} (\Theta, \phi)) = 2 \pi \delta_{kl}
\left\{ \begin{array}{lcl}
1 - \frac{1}{2} \sin^2 \Theta & , & k = 1, 2\\
1 - \cos^2 \Theta & , & k = 3 \end{array} \right. \hk .
\ee
Putting the 3-axis of $\vec{q}$-space 
parallel to the vector $\vec{y} - \vec{z}$ we find for
the relevant integral 
\be
\label{26}
\int d \hat{\Omega}_q t_{k l} (\hat{q}) e^{i \alpha \cos \Theta}
 =  2 \pi \delta_{k l} \il^1_{- 1} d z e^{i \alpha z} \left\{ \begin{array}{l}
\frac{1}{2} (1 + z^2) \\
1 - z^2 \end{array} \right. 
 =  2 \pi \delta_{k l} \left\{ \begin{array}{lcl} j_0 (\alpha) - j''_0
 (\alpha) &
, & k = 1, 2 \\
2 (j_0 (\alpha) + j''_0 (\alpha)) & , & k = 3 \end{array} \right.\,
\ee
where $\alpha = q | \vy - \vz |$ and 
we have introduced the spherical Bessel function
$j_0 (\alpha)  = \frac{\sin
\alpha}{\alpha}$.
\bi

\no
To proceed further with the evaluation of the geometric function $H (C, q)$
(\ref{23}) we have to adopt an explicit realization of the loop $C$. For
simplicity we will choose a planar circular loop of radius $R$, 
which we put for
convenience in the 1-2-plane. Then it is convenient to use cylindrical coordinates
in $\vec{x}$-space and to parameterize the loop by the azimuthal
angle $\varphi$
\be
\label{28}
\vec{x} ( \varphi ) = R \vec{e}_\rho (\varphi) \hk , \hk \vec{e}_\rho (\varphi)
= \vec{e}_1 \cos \varphi + \vec{e}_2 \sin \varphi \hk , \hk d \vec{x} = R
\vec{e}_\varphi (\varphi) d \varphi \hk , \hk \vec{e}_\varphi = -
\partial_\varphi \vec{e}_\rho (\varphi) \hk .
\ee
With eq.\ (\ref{26}) the geometric function (\ref{23}) then becomes  
\be
\label{29}
H (C, q) & = & \frac{1}{2} R^2 \il^{2 \pi}_0 d \varphi \il^{2 \pi}_0 d \varphi'
\vec{e}_\varphi (\varphi) \cdot \vec{e}_\varphi (\varphi') f \lk R q |
\vec{e}_\rho (\varphi) - \vec{e}_\rho (\varphi') | \rk \hk ,
\ee
where we have
introduced the abbreviation
\be
\label{30}
f (\alpha) = \frac{1}{2} \il^1_{- 1} d z (1 + z^2) e^{i \alpha z} = \frac{1}{2}
\il^1_{- 1} d z (1 + z^2) \cos (\alpha z) = j_0 (\alpha) -  j''_0 (\alpha) \hk .
\ee
\bi

\no
Using
\be
j'_0 (x) = - j_1 (x) \hk , \hk 3 j'_1 (x) = j_0 (x) - 2 j_2 (x)
\ee
the function $f (x)$ (\ref{30}) can be expressed as
\be
\label{7-36}
f (x) = \frac{2}{3} \lk 2 j_0 (x) - j_2 (x) \rk \hk .
\ee
Since $j_0 (0) = 1$ and $j_{l > 0} (0) = 0$ we have $f (0) = \frac{4}{3}$ and
the integrand in the angular integral is regular everywhere. Note also, that $j_0
(x)$,  $j''_0 (x)$ and $f (x)$ (\ref{30}) are even functions
of $x$.
\bi

\begin{figure}[h]
\includegraphics[scale=\ingrscale]{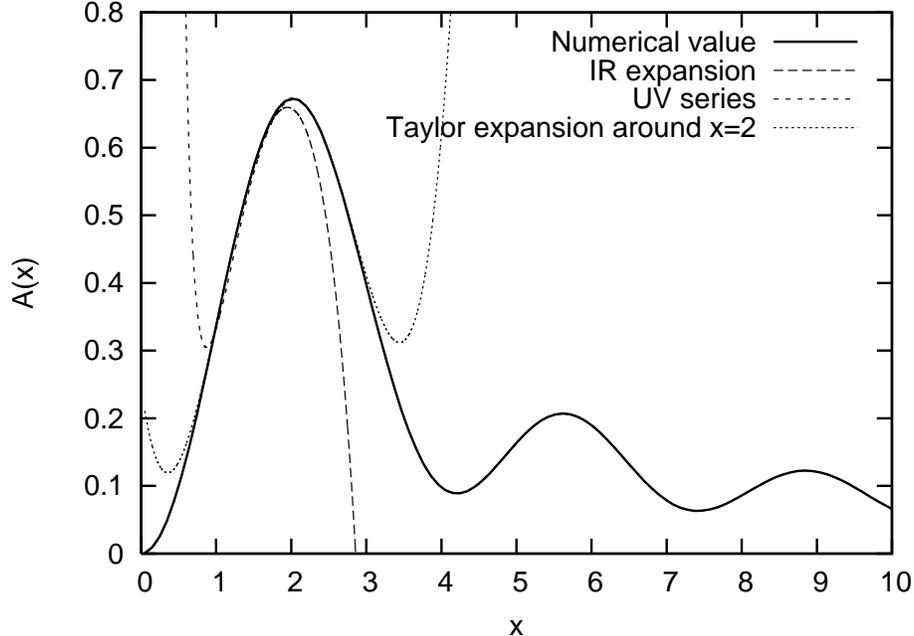}  
\caption{The angular function $A (x)$ (\ref{8-39}): the full numerical result
(solid line), its infrared (dashed line) and ultraviolet (short dashed line)
asymptotic behaviors defined by eqs.\ (\ref{test}) and (\ref{12-48}) and its
Taylor expansion around $x = 2$ (dotted line). See text for more details.}
\label{fig1X}
\end{figure}

\no
The expression (\ref{29}) can be further simplified by using 
\be
\label{7-33}
\vec{e}_\varphi (\varphi) \cdot \vec{e}_\varphi (\varphi') & = & \cos (\varphi -
\varphi') \hk , \hk \left| \vec{e}_\rho (\varphi) - \vec{e}_\rho (\varphi')
\right| = \sqrt{ 2 \lk 1 - \cos (\varphi - \varphi') \rk} \hk 
\ee
resulting in
\be
\label{AA1}
H (C, q) = \frac{1}{2} R^2 \il^{2 \pi}_0 d \varphi \il^{2 \pi}_0 d \varphi' \cos
(\varphi - \varphi') f \lk R q \sqrt{2 (1 - \cos (\varphi - \varphi')} \rk \hk .
\ee
Hence the integrand in (\ref{29}) depends only on the difference $\varphi -
\varphi'$. Therefore we expect that one of the angular integrals can be trivially
taken. This, in fact, turns out to be the case, although the integrations run over
finite intervals. In Appendix \ref{sec-app-a} we show that the geometric factor (\ref{29})
can indeed be reduced to
\be
\label{8-38}
H (C, q) 
& = & 4 \pi R^2 A (R q) \hk , \hk A (x) = 
 \il^{\frac{\pi}{2}}_0  d \alpha (1 - 2 \sin^2 \alpha) f (2 x \sin
\alpha) \hk ,
\ee
Let us stress, 
that the function $H (C, q)$ reflects only the geometry of the considered
loop $C$ and does not contain any dynamical information as is clear from its
definition given in eq.\ (\ref{23}). It contains only the kinemetical information
about the transversality of the gauge field due the presence of the transversal
projector. Note also, 
that the integrand in eq.\ (\ref{8-38}) is well-defined in the
whole integration interval, 
in particular, the function $f (x)$ is well-defined at
the origin. 
After a
sequence of manipulations (see Appendix \ref{sec-app-a}) this integral (\ref{8-38}) can be done analytically with the
result $(x=qR)$
\be
\label{8-39}
A (x) & = & \frac{\pi}{4 x} \left[ 2 x J_0 (2 x) + \pi x \lk J_1 (2 x) H_0 (2 x)
- J_0 (2 x) H_1 (2 x) \rk - 2 J_1 (2 x) \right] \nonumber\\
& & - \frac{3 \pi^2}{16 x^2} \left[ J_2 (2 x) H_1 (2 x) - J_1 (2 x) H_2 (2 x)
\right] \hk ,
\ee
where $J_\nu (x)$ and $H_\nu (x)$ are the ordinary Bessel functions
and the Struve function, respectively.  The asymptotic forms of the
Bessel and Struve functions for small and large arguments are known.
 This yields the following asymptotic behaviors of the angular
integral $A (x)$ for small $x$
\be
\label{test}
A (x) = \frac{2 \pi}{15} x^2 - \frac{\pi}{35} x^4 + \cdots \hk , \hk x \to 0 
\ee
and for large $x$
\be
\label{12-48}
A (x) = \frac{\pi}{4 x} + \frac{\sqrt{\pi}}{2}  \cos \lk 2 x + \frac{\pi}{4} \rk
\frac{1}{\sqrt{x^3}} + O \lk \frac{1}{\sqrt{x^5}} \rk \hk , \hk x \to \infty.
\ee
Fig.\ \ref{fig1X} shows the full function $A (x)$ together with its asymptotic forms.
The asymptotic forms provide excellent approximations in the small
and large momentum region. In the intermediate momentum range $1.5
\leq x \leq 2.5$ we Taylor expand $A (x)$ around $x = 2$ up to
fourth order. Matching the Taylor expansion at $x = 1.5$ and $x =
2.5$, respectively, with the infrared and ultraviolet asymptotic
forms yields a very accurate representation of the function $A (x)$
in the whole momentum range, see fig.\ \ref{fig1X}. This representation is used in the
numerical calculations.

Inserting eq.\ (\ref{8-38}) into eq.\ (\ref{22}) we find for the exponent 
of the 't
Hooft loop
\be
\label{12X}
S (R) = 4 \pi R^2 \il^\infty_0 d q K (q) A (R q) \hk .
\ee
We are interested in the large $R$ behavior of ${S} (R)$. For this purpose
we study first the UV- and IR-behavior of the integrand, in particular of the
kernel $K (q)$. This requires the asymptotic behaviors of the solutions of the
Dyson-Schwinger equations.

\section{The Dyson-Schwinger equations revisited}\label{sec-revisited}

\subsection{Ultraviolet behavior and renormalization}
Consider the (unrenormalized) gap equation obtained in ref. \cite{Feuchter:2004mk} by  
variation
of the energy density with respect to the kernel $\omega$
\be
\label{re1}
\omega^2 (k) = k^2 + \chi^2 (k) + I_\omega (k) + I^0_\omega \hk ,
\ee
where
\begin{align}
\label{re2x}
\chi ({\mybf k})  &\equiv I_\chi (k) = \frac{N_C}{4} \int \frac{d^3 q}{(2 \pi)^3}  
\lk 1 - (\hat{{\mybf k}}
\hat{{\mybf q}})^2 \rk \frac{d ({\mybf k} -{\mybf q}) d ({\mybf q})}{({\mybf k} - {\mybf q})^2}\\
\label{re2}
I_\omega ({\mybf k}) &= \frac{N_C}{4} \int \frac{d^3 q}{(2 \pi)^3} \lk 1 + (\hat{{\mybf k}} \hat{{\mybf q}})^2
\rk 
 \cdot \frac{d ({\mybf k} -{\mybf q})^2 f ({\mybf k} - {\mybf q})}{({\mybf k} - {\mybf q})^2} 
\cdot \frac{\left[ \omega ({\mybf q}) - \chi ({\mybf q}) + \chi
({\mybf k}) \right]^2 - \omega ({\mybf k})^2}{\omega ({\mybf q})} \\
\label{re2a}
I^0_\omega &= \frac{N_C}{4} g^2 \int 
\frac{d^3 q}{(2 \pi)^3}
\lk 3 - (\hat{{\mybf k}} \hat{{\mybf q}})^2 \rk \frac{1}{\omega ({\mybf
q})}
\end{align}
are all UV-divergent integrals, so that this equation needs regularization and
renormalization. In the above integrals $d (k)$ and $f (k)$ are, respectively,
the ghost (see eq.\ (\ref{ghost-propagator})) and the Coulomb form factor, which are determined by two  
further DSEs,
whose renormalization is described in detail in ref. \cite{Feuchter:2004mk}.
\bi

\no
We renormalize the gap equation (\ref{re1}) in the standard way
by subtracting it at an arbitrary
renormalization point $\mu$. Defining
\be
\label{re4}
\bar{\chi} (k) = \chi (k) - \chi (\mu) \equiv I_\chi (k) - I_\chi (\mu)
\ee
and
using $\bar{\chi} (\mu) = 0$ we obtain
\be
\label{re5}
\omega^2 (k) = k^2 + \bar{\chi}^2 (k) + \omega^2 (\mu) - \mu^2 + 2
\bar{\chi} (k)
\chi (\mu) + I_\omega (k) - I_\omega (\mu) \hk .
\ee
The renormalized curvature $\bar\chi(k)$ (\ref{re4}) is UV-finite.
The crucial observation now is that in
$I_\omega (k)$ (\ref{re2}) the UV-divergent
piece $\chi (\mu)$ drops out. We can hence replace $\chi (k)$ in  
$I_\omega (k)$
(\ref{re2}) by $\bar{\chi} (k)$. Analogously to
  ref. \cite{Feuchter:2004mk} we rewrite the
Coulomb integral $I_\omega (k)$ (\ref{re2}) as
\be
\label{re6}
I_\omega (k) = I^{(2)}_\omega[\bar\chi] (k) + 2 \bar{\chi} (k) I^{(1)}_\omega[\bar\chi] (k) \hk ,
\ee
where the integrals
\begin{align}
\label{re7}
I^{(n)}_\omega[\chi] (k) &= \frac{N_C}{4} \int \frac{d^3 q}{(2 \pi)^3} \lk 1 
+ (\hat{{\mybf k}} \hat{{\mybf q}})^2 \rk  
\cdot \frac{d ({\mybf k} -{\mybf q})^2 f ({\mybf k} - {\mybf q})}{({\mybf k} - {\mybf q})^2} 
\cdot \frac{\left[\omega ({\mybf q}) - \chi ({\mybf q})\right]^n - \left[\omega ({\mybf k}) 
- \chi ({\mybf k})\right]^n }
{\omega ({\mybf q})}
\end{align}
are for $n = 1$ and $n = 2$, respectively,
  linearly and quadratically ultraviolet
divergent, while the differences
\be
\label{re8}
\Delta I^{(n)}_\omega (k) = I^{(n)}_\omega[\bar\chi] (k) - I^{(n)}_\omega[\bar\chi] (\mu)
\ee
are ultraviolet finite. From eq. (\ref{re6}) we
obtain with $\bar{\chi} (\mu) = 0$
\be
\label{re9}
I_\omega (k) - I_\omega (\mu) = \Delta I_\omega ^{(2)} (k) + 2 \bar{\chi} (k)
I^{(1)}_\omega (k) \hk .
\ee
With this result the renormalized gap equation (\ref{re5}) can be  
rewritten as
\be
\label{re10}
\omega^2 (k) - \bar{\chi}^2 (k) =
k^2 + \xi_0 + \Delta I^{(2)}_\omega (k)
+ 2 \bar{\chi} (k) \lk \xi + \Delta I^{(1)}_\omega
  (k) - \Delta
I^{(1)}_\omega (0) \rk  \hk ,
\ee
where we have introduced the abbreviations
\be
\label{re11}
\xi_0 & = & \omega^2 (\mu) - \mu^2
\\
\label{re12}
\xi & = & \chi (\mu) + I^{(1)}_\omega (0) \hk .
\ee
Since $\omega (\mu)$ and $\mu$ are finite constants%
, $\xi_0$ is a finite renormalization  
constant. The
only singular piece in the gap equation (\ref{re10}) is the quantity $\xi$
(\ref{re12}). Ignoring the anomalous dimensions
both $\chi (\mu)$ and $I^{(1)}_\omega (0)$ are linearly UV-divergent.
These linearly diverging terms of the 3-dimensional canonical Hamilton  
approach
correspond to quadratically divergent terms in the 4-dimensional covariant
Lagrange formulation. As is well known these singularities have to
  cancel by gauge
invariance. We should hence
ignore these singularities (which are artifacts due to our approximations)
and keep only the finite parts of $\chi(\mu)$ and $I_\omega^{(1)}(0)$. In the following we therefore consider $\chi (\mu), I^{(1)}_\omega (0)$ and thus
$\xi$ as finite renormalization constants.
\medskip

\no
The novel result of the above consideration is that in the gap equation there
are only two independent renormalization constants, $\xi_0$ and $\xi$.
The renormalized gap equation obtained in ref. \cite{Feuchter:2004mk} seems to depend  
explicitly on the
three renormalization constants: $\xi, \xi_0$ and $\chi (\mu)$. The constant
$\chi (\mu)$ sneaked  formally into the renormalized gap equation  
since it was
not realized that this quantity drops out from the Coulomb integral $I_\omega(k)$ (\ref{re2}). However, it was already found empirically in ref. \cite{Feuchter:2004mk} that the numerical solutions to the DSE are practically independent of $\chi (\mu)$. Indeed, by noticing that the integrals $I_\omega^{(n)}[\chi](k)$ defined by eq.\ (\ref{re7}) satisfy the relations
\begin{align}
I_\omega^{(1)}[\bar\chi+\chi(\mu)](k) &= I_\omega^{(1)}[\bar\chi](k) \\
I_\omega^{(2)}[\bar\chi+\chi(\mu)](k) &= I_\omega^{(2)}[\bar\chi](k)-2\chi(\mu)I_\omega^{(1)}[\bar\chi](k) 
\end{align}
one shows that the renormalized gap equation (\ref{re10}) agrees with the one derived in ref.\ \cite{Feuchter:2004mk}. Thus
all results derived in ref.\ \cite{Feuchter:2004mk} about the behavior of the  
solutions of the
Dyson-Schwinger equations remain true. This refers in particular to the fact that
both, the infrared and the
ultraviolet behavior of the solutions of the gap equation are insensitive to
the precise value of the renormalization constants
(except for the renormalization
constant of the ghost form factor which is fixed, however, by the horizon
condition, see below).
While the (renormalized) gap equation obtained by minimizing the  
energy density
does not depend on the choice of $\chi (\mu)$ (but merely on the  
values of $\xi$
and $\xi_0$) other observables may explicitly depend on $\chi (\mu)$. In fact,
we will see that the 't Hooft loop does depend on $\chi (\mu)$.
\bi

\no

The DSE resulting from the minimization of the energy
density can be solved analytically not only in the ultraviolet where
perturbation theory is applicable but also in the infrared
\cite{Feuchter:2004mk,Schleifenbaum:2006bq}. 
\bi

\no
\subsection{Infrared behavior}
\bi

\no
In ref.\ \cite{Feuchter:2004mk} it was shown, that the infrared behavior of the solutions of
the DSE is rather uniquely determined once the so-called
``horizon condition'', \cite{Zwanziger:1998ez}, 
i.e.\ an infrared divergent ghost form factor, $d^{- 1} (k=0) = 0$,
is implemented. This condition turns out to be absolutely necessary in $D = 2 +
1$-dimensions to obtain a self-consistent solution to the Dyson-Schwinger
equation in Coulomb gauge \cite{R10}. In $D = 3 + 1$ this condition is in
accord with the Gribov-Zwanziger confinement scenario, which is consistent with 
the Kugo-Ojima confinement criteria in Landau gauge \cite{Kugo:1979gm}. 
\bi

\no
Infrared  analysis of the ghost (or gluon) Dyson-Schwinger equation (in the
rainbow-ladder approximation), implementing the horizon condition $d^{- 1} (k =
0) = 0$ and using the power law ans\"atze
\be
\label{13-51}
\omega (k) = \frac{a}{k^\alpha} \hk , \hk d (k) = \frac{b}{k^\beta}
\ee
yields the sum rule (in $D=d+1$ dimensions) \cite{Schleifenbaum:2006bq}
\be
\alpha = 2 \beta + 2 - d 
\ee
due to the non-renormalization of the ghost-gluon vertex. This sum rule also
guarantees that $\chi (k)$ (\ref{re2x}) (and thus $\bar\chi(k)$ (\ref{re4})) has the same infrared power as $\omega (k)$,
i.e.\
\be
\label{12-45}
\bar\chi (k) = \frac{u}{k^\alpha} \hk , \hk k \to 0 \hk .
\ee
The horizon condition implies $\beta > 0$ and for $\beta > \frac{d-2}{2}$ the gluon
energy $\omega (k)$ (and thus also $\bar\chi (k)$) is infrared divergent,
i.e.\ $\alpha > 0$. It can be shown that the gap equation has indeed a consistent solution where both $\omega(k)$ and $\bar\chi(k)$ are infrared divergent and in addition their difference $(\omega(k)-\bar\chi(k))$ is infrared finite.

Indeed if $(\omega(k)-\bar\chi(k))$ is infrared finite it follows that $I_\omega^{(n)}(k)$ and thus $\Delta I_\omega^{(n)}(k)$ are infrared finite. Then for
infrared divergent $\omega(k)$ and $\bar\chi(k)$ the gap equation (\ref{re10}) reduces for $k \to 0$ to
\begin{align}
\label{62}
\left(\omega(k)+\bar\chi(k)\right)\left(\omega(k)-\bar\chi(k)\right)=2\xi\bar\chi(k)\, .
\end{align}
Since $\omega(k)$ and $\bar\chi(k)$ have the same infrared exponent it follows from (\ref{62}) that the infrared limit of the gap equation is given by
\be
\label{12-46}
\lim\limits_{k \to 0} \lk \omega (k) - \bar\chi (k) \rk = \xi
\ee
or expressed in terms of the total curvature $\chi(k)=\bar\chi(k)+\chi(\mu)$
\begin{align}
\label{64}
\lim\limits_{k\to 0} (\omega(k)-\chi(k)) = c\, ,\qquad c \equiv \xi - \chi(\mu)
\end{align}
Instead of $\chi(\mu)$ we can use $c$ as independent renormalization constant.
Equation (\ref{12-46}) implies (c.f.\ eqs.\ (\ref{13-51}) and (\ref{12-45}))
\begin{align}
\label{12-47}
u = a \hk .
\end{align}
Equation (\ref{12-46}) along with the
ghost DSE in the infrared limit can be solved analytically for the infrared
exponents yielding two solutions
\be
\beta \approx 0.796 \hk , \hk \beta = 1 \hk .
\ee
Only the latter one gives rise to a strictly linear rising static
quark potential at $r \to \infty$, 
\be
V (r) = \frac{1}{2 \pi^2} \il^\infty_0 d k (d (k))^2 \left[ 1 - \frac{\sin k
r}{k r} \right] \hk .
\ee
Here we have assumed that the so-called Coulomb form factor can be put to one
\cite{Feuchter:2004mk}.
The first term in the bracket represents the self-energies of the static (quark and antiquark) 
color charges. In this paper we will use the solution $\beta = 1$ because the resulting linear rise of the static potential allows us to fix the scale by the (Coulomb) string tension.
\bi

\no
By definition the (Coulomb) string tension is given by that part of
\be
\label{12X2}
\frac{d V}{d r}
 = - \frac{1}{2 \pi^2} \il^\infty_0 d k k (d (k))^2 \lk \frac{d}{d x}
\frac{\sin x}{x} \rk_{x = k r} \hk ,
\ee
which (for large $r$)
is independent of $r$. This part is exclusively determined by the infrared
behavior of $d (k)$. 
Inserting the infrared form (\ref{13-51}) (with $\beta = 1$)  into eq.\
(\ref{12X2}) we obtain for the Coulomb string tension
\be
\label{eq-coulomb-string-tension}
\sigma_C = - \frac{b^2}{2 \pi^2} \il^\infty_0 \frac{d x}{x} \frac{d}{d x} \lk
\frac{\sin x}{x} \rk \hk .
\ee
This integral can be done, yielding
\be
\sigma_C = \frac{b^2}{8 \pi} \hk .
\ee
Thus the infrared coefficient of the ghost form factor can be explicitly
expressed by the string tension. The infrared analysis of the 
ghost Dyson-Schwinger equation
provides the relation between the infrared coefficients in (\ref{13-51}) 
\cite{Schleifenbaum:2006bq}
\be
a = \frac{b^2 N_c}{4 (4 \pi){\raisebox{3mm}{\tiny d} \hspace{ -1mm }
\raisebox{2mm}{\tiny /} \hspace{ -1mm }
\raisebox{1mm}{\tiny 2} }} \frac{\Gamma \lk \frac{d - \beta}{2} \rk^2 \Gamma \lk
1 - \frac{d}{2} + \beta \rk}{\Gamma (d - \beta) \Gamma \lk 1 + \frac{\beta}{2}
\rk^2} \hk .
\ee 
Hence the infrared coefficient $a$ of $\omega(k)$ and $\chi (k)$ (see eqs.\
(\ref{13-51}), (\ref{12-45}), (\ref{12-47})) can be also expressed
by the string tension.
For $d = 3$ and $\beta = 1$ the last relation simplifies to
\be
a = b^2 \frac{N_c}{8 \pi^2} \hk .
\ee
Lattice calculations \cite{Greensite:2004ke}, \cite{Langfeld:2004qs}
show that $\sigma_C$ is by a factor of 1.5 -- 3 larger
than the string tension $\sigma$ extracted from the 
Wilson loop, in accord with the fact that
$\sigma_C$ gives an upper bound to $\sigma$
\cite{Zwanziger:2002sh}.  Assuming a value of the
Coulomb string tension $\sigma_C = 1.5 \sigma (\sigma = (440 MeV)^2)$ we find
\be
a = 0.185\,\mathrm{GeV}^2, \hspace{1cm} b = 2.702\,\mathrm{GeV} \hk .
\ee
These values are in remarkable agreement with the figures extracted from the
numerical solutions of the Dyson-Schwinger equations.

\section{The asymptotic behavior of the 't~Hooft loop}\label{sec-asymptotic} 

Given that $\omega (k)$ and $\chi (k)$ are both infrared singular and satisfy
the relation (\ref{64}) it follows that the infrared limit of the kernel $
K (k)$ (\ref{6-24}) is given by
\be 
\label{39}
\lim\limits_{k \to 0} K (k) = \lim\limits_{k \to 0} (\omega (k) - \chi (k)) =
c \hk .
\ee

\no
As we will explicitly see further below, it is
 the infrared behavior of $K (k)$ which
determines the large $R$-behavior of the 't~Hooft loop. The fact that $K (q)$
is infrared finite (see eq.\ (\ref{39})) restricts the possible $R$-dependence of
$S$ (\ref{12X}) drastically.

\no
\subsection{Renormalization of the 't~Hooft loop}
\bi

\begin{figure}[h]
\includegraphics[scale=\ingrscale]{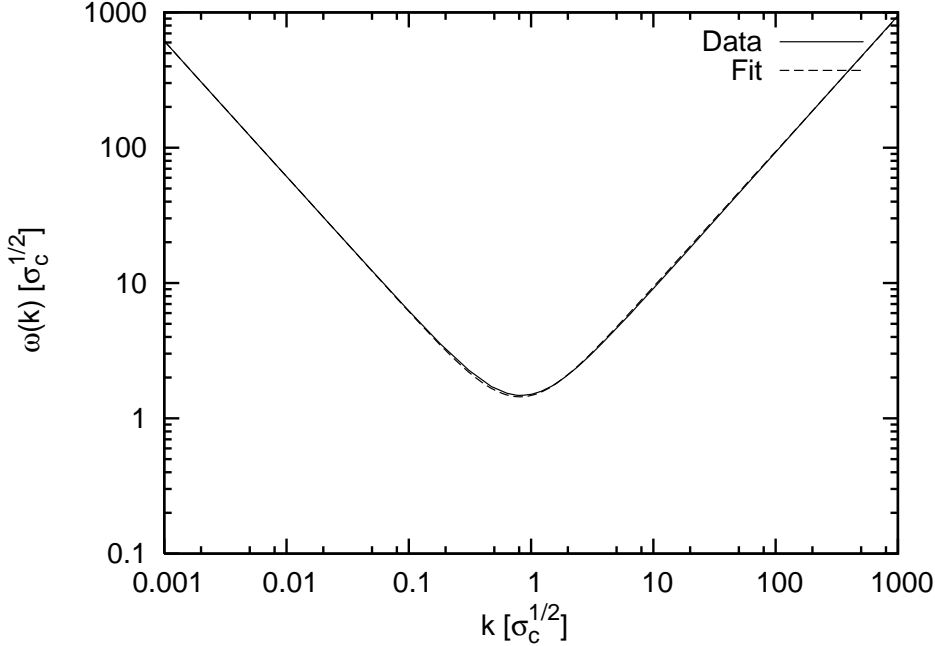}
\caption{The gluon energy resulting from the numerical solution of the
Dyson-Schwinger equation \cite{Epple:2006hv} 
and its parameterized form (\ref{15-63}).}
\label{fig2X}
\end{figure}

\no
In ref.\ \cite{Feuchter:2004mk} it was shown that in the ultraviolet $(k \to
\infty)$
\be
\label{14-55}
\omega (k) \to k \hk , \hk \frac{\chi (k)}{\omega (k)} \to 
\frac{{c'}}{\sqrt{\ln
\frac{k}{\mu}}} \hk ,
\ee
where $\mu$ is some arbitrary energy scale (renormalization point) and ${c'}$
 is
some constant whose precise value depends on the choice of $\mu$. 
The first relation expresses asymptotic freedom, i.e.\ for
large momenta gluons become photons, while the second relation expresses the presence of an
anomalous dimension. With (\ref{14-55}) 
for large $q \to \infty$  the kernel $K (q)$ defined by eq.\
(\ref{6-24}) behaves asymptotically  as 
$K (q) \to \frac{1}{2} q$. 
Since $A
(x = q R) \sim \frac{1}{x}$ for $x \to \infty$, see eq.\ (\ref{12-48}), 
it follows that the exponent of the 't~Hooft loop $S$ (\ref{12X}) is linearly
UV-divergent. UV-singularities are expected given the singular nature of the 't
Hooft loop. There are also UV-subleading terms in $K (q)$ which give rise to
UV-singularities in $S$ (\ref{12X}). To extract these terms let us parameterize
$\omega (q)$ by
\be
\label{15-63}
\omega (k) = \frac{a}{k} + a_0 + k \hk ,
\ee
which is in accord with the asymptotic IR- and UV-behavior and 
which gives a good
approximation to the numerical solution of the gap 
equation in the whole momentum
range (see fig.\ \ref{fig2X}). For the solution presented in fig.\ \ref{fig2X} we find numerically $a = 0.61$ and $a_0 = -0.078$.

\bi

\no
The leading UV-terms of $K (q)$ giving rise to UV singularities in $S$
(\ref{12X}) are then given by
\be
K (q) & \to & \frac{1}{2} (q + a_0) \lk 1 - \frac{\bar{c}}{\ln \frac{q}{\mu}} +
\frac{c_{- 1}}{q} + \dots \rk \nonumber\\
& = & \frac{1}{2} q - \frac{1}{2} q \frac{\bar{c}}{\ln \frac{q}{\mu}} +
\frac{1}{2} \lk a_0 + c_{- 1} \rk - a_0 \frac{\bar{c}}{\ln \frac{q}{\mu}}  +
\dots \hk .
\ee
with unknown coefficicents $\bar{c}$, $c_{- 1}$ arising from the UV-expansion of $\chi(q)$.
These terms produce UV-singularities of $S$ (\ref{12X}) of the form ($\Lambda$ -
UV-cut-off)
\be
\sim \Lambda \hk , \hk \mathrm{li} (\Lambda) \hk  , \hk \ln \Lambda \hk , \hk \ln (\ln
\Lambda) \hk ,
\ee
where $\mathrm{li} (x)$ is the logarithmic integral, which behaves for large $x$ as
$\mathrm{li} (x) = \pi (x) + O \lk \sqrt{x} \ln x \rk$, 
where $\pi (x)$ is the number of primes less than or equal to $x$.
In practice it is not possible to determine the various coefficients of the
UV-asymptotic expansion of $\chi (q)$ with sufficient accuracy. We therefore
propose to eliminate the UV-divergencies of $S$ (\ref{12X}) by subtracting from
the kernel $K (q)$ (\ref{6-24}) the following UV-leading one
\be
\label{14-60}
K_0 (q) = \frac{1}{2} (q + a_0) \lk 1 - \lk \frac{\chi (q)}{\omega (q)} \rk^2 
\rk \hk ,
\ee
which results from $K (q)$ (\ref{6-24}) by replacing $\omega (q)$ by its 
UV-leading and next to leading order terms (see eq.\ (\ref{15-63})), while keeping
the ratio $\chi (q) / \omega (q)$ fixed.
The kernel $K_0 (q)$ (\ref{14-60}) 
contains all the UV-leading terms which give rise to
UV-singularities in $S$. In addition (since we have kept fully $\lk 1 - \lk
\frac{\chi}{\omega} \rk^2 \rk $) it also contains terms which produce finite
contributions to $S$. However, these terms are negligible compared to the
dominating finite terms of $S$ produced by the full $K$. The reason is the
following:
For large $R$ the 't~Hooft loop is dominated by the infrared part of $K (q)$ as
one explicitly notices by rescaling the integration variable $q \to q R = x$ in
(\ref{12X}) and as we will explicitly show further below. Furthermore, since
$\omega (q) \sim 1/q$ for $q \to 0$, in the infrared $K_0 (q)$ is by a power of
$q$ suppressed relative to $K (q)$
 and thus gives sub-leading contributions to the large 
$R$ behavior of 
the 't
Hooft loop. By the same token, elimination of the UV-divergent terms from $S$ by 
replacing $K (q)$ by  $K (q) - K_0 (q)$ does not change the large $R$-behavior of
$S$.
\bi

\no
 We
thus
isolate the ultraviolet 
divergent parts of ${S}$ (\ref{12X}) by writing
\be
S = (S - S_0) + S_0 \hk ,
\ee
where $S_0$ results from $S$ (\ref{12X}) by replacing $K (q)$ (\ref{6-24}) 
by $K_0 (q)$ (\ref{14-60}). 
By construction $S - S_0$ is UV-finite. The leading UV-divergent term 
$S_0^{(l)}$ of
 $S_0$ is independent of $R$ and $\exp (- S^{(l)}_0) = Z$ can 
 be absorbed into the 
renormalization of the wave functional $Z \raisebox{3mm}{\scriptsize 1}
 \hspace{ -0,9mm }
\raisebox{2mm}{\scriptsize /} \hspace{ -0,9mm }
\raisebox{1mm}{\scriptsize 2}
\Psi \to \Psi$. The remaining
terms in $S_0$ have weak $R$-dependences (at most
 logarithmic) and as explained
above their elimination does not spoil the use of the renormalized 't~Hooft loop
\be
\label{16XX}
\bar{S} = S - S_0 = 4 \pi R \il^\infty_0 d x \lk K \lk \frac{x}{R} \rk - K_0 
\lk \frac{x}{R} \rk \rk A (x)
\ee
as a disorder parameter, i.e.\ they
do not convert a perimeter law into an area law or vice versa. Eq.\ (\ref{16XX})
explicitly shows that the large $R$-behavior of the 't~Hooft loop is determined by the 
infrared behavior of $K (q)$.

\no

\subsection{The large $R$ behavior of the 't~Hooft loop}
\bi

\no
We are interested in the 't~Hooft loop, i.e.\ in the quantity $\bar{S} (R)$
defined by eq.\ (\ref{16XX}) for large $R \to \infty$. If we could take in eq.\
(\ref{16XX}) the limit $R \to \infty$ before taking the momentum integral, using
\be
\lim\limits_{R \to \infty} K \lk \frac{x}{R} \rk & = & K (0) = c \nonumber\\
\lim\limits_{R \to \infty} K_0 \lk \frac{x}{R} \rk & = & K_0 (0) = 0 \hk ,
\ee
we would obtain a perimeter law
\be
\label{16-63}
\bar{S} = R \kappa \hk , \hk \kappa = 4 \pi c \il^\infty_0 d x A (x) 
\ee
for $c \neq 0$. However, the remaining integral (\ref{16-63}) 
is UV-divergent while the
original integral (\ref{16XX})  is convergent. Thus we must not take the 
limit
$R \to \infty$ before having carried out the momentum integral: In fact, we will 
show
below that a finite $c \neq 0$ gives rise to a large $R$-behavior $\bar{S} (R)
\sim R \ln R$, while $c = 0$ yields a perimeter law $\bar{S} (R) \sim R$.
\bi

\no
The fact that we must not take the limit $R \to \infty$ in (\ref{16XX}) before
taking the integral does not, however, imply that $\bar{S}$ is not dominated by
the infrared part of $K (q)$. We will explicitly show further below that it is
the infrared part of the momentum integral (\ref{16XX}) (and thus of $K (q)$)
which determines whether an area law or perimeter law arises.
\bi

\no
Since $\omega (k) - \chi (k)$ is infrared finite (by the gap equation, see eq.\ (\ref{12-46})) it is of
the form
\be
\label{17-67}
\omega (k) - \chi (k) =\bar{C} k^\gamma \hk , \hk \gamma \geq 0 \hk , \hk k \to 0 \hk,
\ee
and by eq.\ (\ref{39}), $K (k)$ has the same IR-behavior\footnote{In the notation of eq.\ (\ref{39}) $\gamma > 0$ corresponds to $c = 0$ and $\gamma = 0$ to $ c
 = \bar{C} \neq 0$.} 
 \be
 K (k) = \bar{C} k^\gamma \hk , \hk \gamma \geq 0 \hk , \hk k \to 0 \hk .
 \ee
Since the large $R$-behavior of $S (R)$ is determined by the infrared
behavior of $K (q)$ and since the UV-behavior of $K (q)$ and $A (R q)$
determines the convergence property of the momentum integral, for the
following qualitative considerations it is sufficient to approximate $
A(x)$ by the interpolating formula
\be
\label{17Y2}
A (x) = \frac{2 \pi}{15} \frac{x^2}{\eta x^3 + 1} \hk , \hk \eta = \frac{8}{15}
\ee
which has the correct IR and UV behavior given in eqs.\ (\ref{test}) and
(\ref{12-48}), respectively. Along the same lines we use the interpolating
formula
\be
\label{17Y1}
\Delta K (q) : = K (q) - K_0 (q) = \bar{C} 
\frac{\lk q / q_0 \rk^\gamma}{(q / q_0)^{\gamma + 1} + 1}
\ee
which has the correct IR and UV behavior. With eqs.\ (\ref{17Y1}) and
(\ref{17Y2}) we find from (\ref{16XX}) 
\be
\label{18-69}
\frac{15}{8 \pi^2 \bar{C} R} \bar{S} = \il^\infty_0 d x \frac{x^\gamma q_0
R}{x^{\gamma + 1} + (q_0 R)^{\gamma + 1} } \frac{x^2}{\eta x^3 + 1} =: I_{IR} +
I_{UV} \hk .
\ee
To estimate this integral we 
split the integration range into the intervals $[0, q_0 R]$ and $[  q_0 R , 
\infty)$ and denote the corresponding integrals by $I_{IR}$ and $I_{UV}$.
Consider first $I_{IR}$ where $0 \leq x \leq q_0 R$. An upper and lower bound to
this integral is obtained by replacing $\frac{1}{x^{\gamma + 1} + (q_0
R)^{\gamma + 1}}$ by $\frac{1}{(q_0 R)^{\gamma + 1}}$
and $\frac{1}{(q_0 R)^{\gamma + 1} + (q_0 R)^{\gamma + 1}} = \frac{1}{2} 
\frac{1}{(q_0 R)^{\gamma + 1}}$, respectively. Thus
we obtain
\be
\label{17Y3}
\frac{1}{2} I^0_{IR} < I_{IR} < I^0_{IR} \hk ,
\ee
where
\be
\label{18-71}
I^0_{IR} & = & \lk \frac{1}{q_0 
R} \rk^\gamma \il^{q_0 R}_{0} d x \frac{x^{\gamma +
2}}{\eta x^3 + 1}
 \hk .
\ee
An upper and lower bound to the integral $I_{UV}$ is obtained by replacing
$\frac{1}{x^{\gamma + 1} + (q_0 R)^{\gamma + 1}}$ by $\frac{1}{x^{\gamma + 1}}$
 and $\frac{1}{x^{\gamma + 1} + x^{\gamma + 1} } = \frac{1}{2 x^{\gamma + 1}}$,
respectively. This yields
\be
\label{17Y4}
\frac{1}{2} I^0_{UV} < I_{UV} < I^0_{UV} \hk ,
\ee
where
\be
I^0_{UV} & = & q_0 R \il^\infty_{q_0 R} d x \frac{x}{\eta x^3 + 1}
\nonumber\\
& = & \frac{1}{3\eta} y \left[ \sqrt{3} \lk \frac{\pi}{2} - \arctan \frac{2 y -
1}{\sqrt{3}} \rk + \frac{1}{2} \ln \frac{y^2 + 2 y + 1}{y^2 - y + 1} \right] \hk
.
\ee
Here we have introduced the abbreviation 
$y=\eta^{1/3}q_0 R$.
 Note that $I^0_{UV}$ is independent of
$\gamma$ and thus independent of the IR-behavior of $\Delta K (q)$. 
For large $R$ the last expression becomes
\be
\label{17Y5}
I^0_{UV} = \frac{1}{\eta} \left[ 1 - \frac{1}{4\eta} \frac{1}{(q_0 R)^3} \right] + 0
\lk \frac{1}{R^5} \rk \hk .
\ee
Thus the UV-part contributes for large $R$ always a perimeter term to $\bar{S}$
(\ref{18-69}). It is the IR-part, $I_{IR}$, or more precisely the infrared exponent
$\gamma$ of $K (q)$, which decides whether the 't~Hooft loop has an area law or
a perimeter law. 
\bi

\no
Consider first $\gamma = 0$. In this case the IR-integral
(\ref{18-71}) becomes
\be
I^0_{IR} = \il^{q R}_0 d x \frac{x^2}{\eta x^3 + 1} = \frac{1}{3\eta} \ln \left[ 1 + \eta (q_0 R)^3 \right]
\ee
and gives rise to an $R$-dependence
\be
\bar{S} \sim R \ln R \hk , \hk R \to \infty \hk .
\ee
Consider now the case $c = 0$, i.e.\ $\gamma > 0$. For $\gamma > 0$ the
IR-integral (\ref{18-71}) can be expressed as
\be
I_{IR} (R) & = & - \pi \left[ 3 \eta (\eta^{1/3} q_0 R)^\gamma \sin \frac{\pi \gamma}{3}
\right]^{- 1} \nonumber\\
& & - \frac{1}{3 \eta} \Phi \lk - \frac{1}{\eta (q_0 R)^3} \hk , \hk 1 \hk , \hk -
\frac{1}{3} \gamma \rk \hk ,
\ee
where
\be
\label{19XX}
\Phi (z, s, \alpha) = \sli^\infty_{k = 0} \frac{z^k}{(\alpha + k)^s}
\ee
is the Lerch transcendent, which is a generalization of the Hurwitz zeta
function and the polylogarithm function and is defined for $|z| < 1$ and $\alpha
\neq 0 , - 1 , \dots$. For large $R$ we can  expand $\Phi (z, 1, -
\frac{1}{3} \gamma)$ in powers of $z = - 1 / (\eta q_0 R)^3$, keeping only the
first two terms in (\ref{19XX}).  In this order we obtain
\be
I_{IR} (R) & = & - \pi \left[ 3 \eta \lk \eta^{1/3} q_0 R \rk^\gamma \sin \frac{\pi
\gamma}{3} \right]^{- 1} \nonumber\\
& & + \frac{1}{\eta\gamma} + \frac{1}{3 \eta} \frac{1}{\eta (q_0 R)^3 ( 1 -
\frac{1}{3} \gamma)} \hk .
\ee
For $\gamma \to 0$ the first two terms have simple poles which cancel. For
$\gamma > 0$ the second term is the leading one for large $R$, and gives rise to
a perimeter contribution to $\bar{S}$ (\ref{18-69}). 
Since the UV integral (\ref{17Y5})
contributes a perimeter term independent of $\gamma$ we obtain for $\gamma > 0$
indeed a perimeter law for the 't~Hooft loop.
\bi

\no
We have thus shown that the Yang-Mills vacuum wave functional obtained in the
variational solution of the Yang-Mills Schr\"odinger equation in Coulomb gauge
\cite{Feuchter:2004mk,Reinhardt:2004mm,Epple:2006hv} yields
indeed a perimeter law for the 't~Hooft loop, provided the renormalization
constant (\ref{39}) is chosen $c = 0$. The 't~Hooft loop is thus more sensitive
to the details of the wave functional than the confinement properties, i.e.\ the
infrared behaviors of the ghost and gluon propagators, which turned out to be
independent of the choice of the renormalization constants once the horizon
condition is implemented \cite{Feuchter:2004mk,Epple:2006hv}. The choice
$c = 0$ is also preferred by the variational principle as we will discuss in
sect. \ref{sec-concl}. 
\bi

\no 
\section{Numerical results}\label{sec-results}
\bi

\begin{figure}[h]
\includegraphics[scale=\ingrscale]{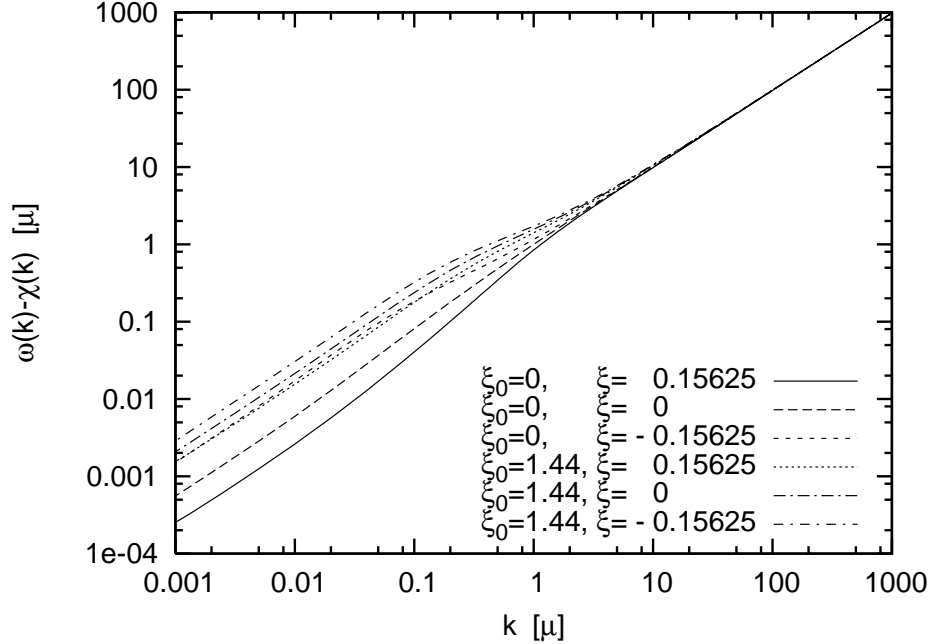}
\caption{The difference $\bar\omega(\bar k)-\bar\chi(\bar k)$ for various choices of the renormalization constants. For all solutions presented here the renormalization constant $c=0$ (see eq.\ (\ref{64})) is chosen, resulting in an infrared-vanishing $\bar\omega(\bar k)-\bar\chi(\bar k)$.}
\label{fig3X}
\end{figure}

\no
The above observed qualitative behavior of the 't~Hooft loop is confirmed by
the numerical calculations.
The DSEs resulting from the variational solution of the Yang-Mills Schr\"odinger
equation in Coulomb gauge derived in ref.\ \cite{Feuchter:2004mk} 
and reviewed in sect.\ \ref{sec-revisited}
are solved numerically as described in ref.\
\cite{Epple:2006hv}. 

Since Yang-Mills theory is a scale-free theory, the DSEs can be entirely expressed in terms of dimensionless quantities defined by rescaling all dimensionful quantities with appropriate powers of a (so far arbitrary) scale $\mu$:
\begin{align}
\bar k &= k/\mu,& \bar\mu&=\mu/\mu=1,& \bar \xi &= \xi/\mu,& \bar\xi_0&=\xi_0/\mu^2,\nonumber\\
\bar\omega(\bar k) &= \omega(\bar k \mu)/\mu,& \bar\chi(\bar k) &= \chi(\bar k \mu)/\mu,&\bar d(\bar k) &= d(\bar k \mu),& \bar f(\bar k) &= f(\bar k \mu).
\label{eq-dimensionless-units}
\end{align}
From the numerical solution of the DSEs we calculate the dimensionless Coulomb string tension $\bar\sigma_c$ (\ref{eq-coulomb-string-tension}), which is related to the physical (Coulomb) string tension $\sigma_c$ by
\begin{align}
\sigma_c = \mu^2\bar\sigma_c.
\end{align}
The last relation fixes the scale $\mu$ once a specific value is assigned to $\sigma_c$. Note that $\bar\sigma_c$ depends, in principle, on the renormalization constants $\bar\xi$, $\bar\xi_0$.

\begin{table}
\begin{center}
\begin{tabular}{@{\extracolsep{5mm}}l | l l l l l l}
$\bar\xi_0$       & 0.0      & 0.0     & 0.0     & 1.44     & 1.44    & 1.44    \\\hline
$\bar\xi$         & -0.15625 & 0       & 0.15625 & -0.15625 & 0       & 0.15625 \\\hline
$\bar\sigma_c\:\:$    & 0.047    & 0.173   & 0.340   & 0.261    & 0.385   & 0.285
\end{tabular}
\end{center}
\caption{The dimensionless renormalization constants $\bar\xi$, $\bar\xi_0$ used in the numerical calculations presented in fig.\ \ref{fig3X}, and the resulting dimensionless Coulomb string tension $\bar\sigma_c$.\label{table-3x}}
\end{table}

Fig.\ \ref{fig3X} shows the difference
$\bar \omega (\bar k) - \bar \chi (\bar k)$ for various choices of the remaining renormalization constants $\bar\xi$, $\bar\xi_0$.
In table \ref{table-3x} we quote the resulting dimensionless Coulomb string tension $\bar\sigma_c$.
For all these solutions the kernel $K (q)$ (\ref{6-24}) 
of the 't~Hooft loop has practically
the same infrared behavior (\ref{39}), (\ref{17-67}) with $\gamma \approx 1$.
With these solutions the quantity $\bar{S} (R)$ (\ref{16XX}) of the 't~Hooft
loop is calculated. Thereby we use for the angular integral $A (x)$ (\ref{8-39}) 
the analytic
representation presented at the end of sect.\ \ref{sec-loop}. In the numerical evaluation of
the momentum integral in $\bar{S}$ (\ref{16XX}) we split the UV part of $A (x)$
(see eq.\ (\ref{12-48})) into an oscillating and non-oscillating part. The
non-oscillating part as well as the infrared part is integrated by the standard
Gauss-Legendre method. The integral over the oscillating part of $A (x)$ is
efficiently carried out using the method of integrating half-waves. 
Fig.\ 4 shows the results
obtained for $\bar{S}(R)/R$ (\ref{16XX}) for $c = 0$ and various values of the remaining
renormalization constants $\xi$, $\xi_0$. 
One observes that indeed for large $R$ the quantity $\bar{S}(R)/R$ approaches a
constant independent of the choice of the remaining renormalization constants while requiring $c = 0$.
Although all these solutions with $c = 0$ yield a
perimeter law for the 't~Hooft loop, i.e.\ $\bar{S}(R)\sim R$, the perimeter tension $\kappa$ defined by
\be
\bar{S} (R \to \infty) \to \kappa R
\ee
depends quite
sensitively on the remaining 
renormalization constants $\xi$, $\xi_0$. In fact these renormalization
constants affect the solutions of the DSE only in the intermediate momentum range
while
both the infrared and ultraviolet behaviors are untouched. This shows that the
perimeter tension is sensitive to the details of the wave functional in the
intermediate momentum range. Therefore it would be very helpful to have lattice
measurements of this quantity. Unfortunately this quantity is very difficult to
measure on the lattice. 
\bi

\no
The range of the undetermined renormalization constant implies quite large
changes of the gluon energy in the intermediate momentum range. If we interprete the lowest-lying glueball as a "two quasi-gluon s-state" and adjust the
minimum in $2\omega (k)$ to the glueball energy of 1.65 GeV and furthermore choose
the Coulomb string tension 1.5 $\sigma$ ($\sigma = (440 MeV)^2$) we find the
perimeter tension around 4.4 GeV.
\bi

\begin{figure}[h]
\includegraphics[scale=\ingrscale]{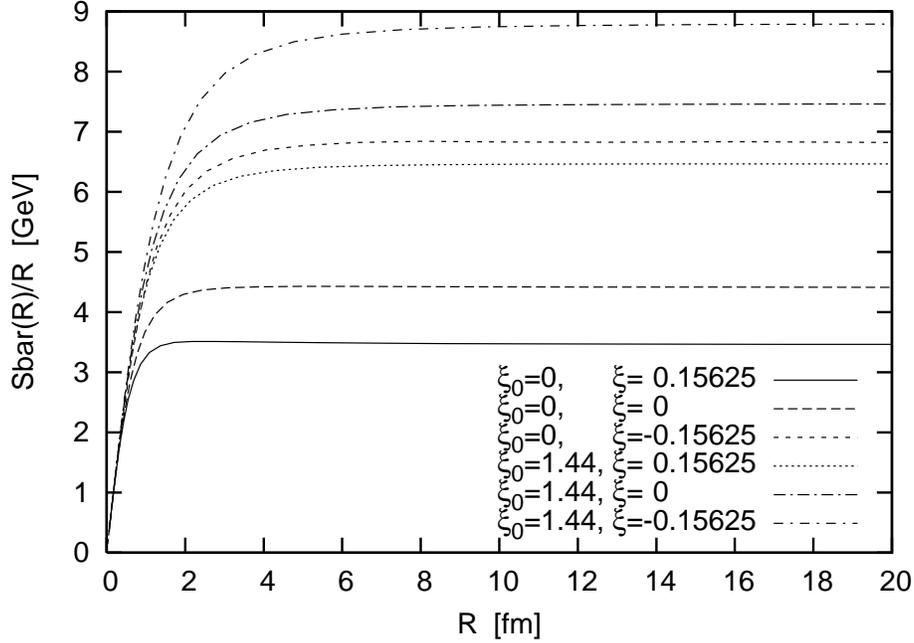}
\caption{The function $\bar{S}(R)/R$, (\ref{16XX}), for different renormalization constants, resulting from the solutions presented in fig.\ \ref{fig3X}.}
\label{fig4X}
\end{figure}


\no
\section{Concluding remarks}\label{sec-concl}
\bi

\no
In summary, the wave functional we have obtained in the variational solution of
the Yang-Mills Schr\"odinger equation yields either a perimeter law $\bar{S}
\sim R$ or $\bar{S} \sim R \ln R$ depending on whether we choose the
renormalization constant (\ref{39}) $c = 0$ or $c \neq 0$, respectively.  The
renormalization constants are undetermined parameters due to the approximations involved in the derivation of the DSEs \cite{Feuchter:2004mk}. However, since
we are using the variational principle the renormalization
constant could, in principle, be considered as variational parameters and fixed by minimizing
the energy density with respect to these parameters. Such type of calculations
are rather expensive, since they require the solution of the DSE in the
multi-dimensional space of the renormalization constants and 
as the energy density
depends only implicitly via the various form factors on these constants. An
exception is the renormalization constant $c$ defined in eq.\
(\ref{64}), which explicitly enters the kinetic energy and which determines the large $R$-behavior of the 't~Hooft loop.
Since $c$ is an infrared property of the
solutions of the DSE for its determination 
it should be sufficient to use the infrared form of these
solutions which are analytically known \cite{Schleifenbaum:2006bq}. 
Then $c$ drops out from the Coulomb energy and does not enter the potential energy, while the kinetic energy density
is proportional to $c^2$ and is minimized by $c = 0$. It is this value which
gives rise to a perimeter law in the 't~Hooft loop. Resorting to the
representation (\ref{12}) (which is correct to two loop in the energy) the value $c
= 0$ implies that our wave functional (\ref{10}) has the form
\be
\Psi (A) = {\cal N} e^{-\frac{1}{2} \int A h A}
\ee
with
\be
h (k) = \omega (k) - \chi (k) \sim k^\gamma \hk , \hk k \to 0 \hk , \hk \gamma >
0 \hk .
\ee
It describes a vacuum where the low momentum modes of the gauge field can fluctuate (for
$k \to 0$) in an unconstrained fashion. Such a stochastic vacuum is
required for an area law in the spatial Wilson loop. We therefore find that a
perimeter law in the 't~Hooft loop is a necessary condition for an area law in the
spatial Wilson loop, which at zero temperature is equivalent to the temporal
one. This is in accord with the duality between the spatial 't~Hooft loop and
the temporal Wilson loop.
\bi

\no
{\Large \bf Acknowledgements}
\bi

\no
Useful discussions with J. Greensite, W. Schleifenbaum, P. Watson and A. Weber
are gratefully acknowledged.
\bi

\no
\begin{appendix}
\section {The loop integral}\label{sec-app-a}
\bi

\no
In the following we work out the geometric factor $H (C, q)$ given by (\ref{AA1}) of the
't~Hooft loop. We first reduce the double integral to a single loop integral,
which eventually can be taken explicitly.
\bi

\no
Since the integrand in eq.\ (\ref{AA1}) 
depends only on the difference $\varphi - \varphi'$ it is
convenient to introduce ``center of mass'' and ``relative'' coordinates
\be
\phi = \frac{\varphi + \varphi'}{2} \hk , \hk \gamma = \varphi - \varphi' \hk .
\ee
The integral (\ref{AA1}) then becomes
\be
H (C, q) = \frac{1}{2} R^2 \left[ \il^\pi_0 d \phi \il^{2 \phi}_{- 2 \phi} d
\gamma + \il^{2 \pi}_{\pi} d \phi \il^{2 (2 \pi - \phi)}_{- 2 (2 \pi - \phi)} d
\gamma \right] \cos \gamma f (R q \sqrt{2 (1 - \cos \gamma)} ) \hk .
\ee 
Changing integration variable in the second term $\phi \to \pi + \phi$ and using
$\cos (- \gamma) = \cos \gamma$ we find
\be
H (C, q) = R^2 \il^\pi_0 d \phi \left[ \il^{2 \phi}_0 d \gamma + \il^{2 (\pi -
\phi)}_0 d \gamma \right] \cos \gamma f (R q \sqrt{2 (1 - \cos \gamma)}) \hk .
\ee
A further change of integration variable $\gamma \to \gamma - 2 \pi$ in the
second term using the
periodicity of $\cos \gamma$, yields
\be
H (C, q) & = & R^2 \il^\pi_0 d \phi \left[ \il^{2 \phi}_0 d \phi \left[ \il^{2
\phi}_0 d \gamma + \il^{- 2 \phi}_{- 2 \pi} d \gamma \right] \right] \cos \gamma f (R q
\sqrt{2 (1 - \cos \gamma)} ) \nonumber\\ 
& = & R^2 \il^\pi_0 \il^{2 \pi}_0 d \gamma \cos \gamma f (q R \sqrt{2 (1 - \cos
\gamma)} ) \nonumber\\
& = & \pi R^2 \il^{2 \pi}_0 d \gamma \cos \gamma f (q R \sqrt{2 (1 - \cos
\gamma)}) \hk .
\ee
As anticipated, one angular integral can be trivially taken. Changing the
integration variable $\alpha = \frac{\gamma}{2}$ we obtain
\be
H (C, q) = 4 \pi R^2 A (R q) \hk ,
\ee
where
\be
\label{AX}
A (x) = \il^{\frac{\pi}{2}}_0 d \alpha (1 - 2 \sin^2 \alpha) f (2 x \sin \alpha)
\hk ,
\ee
which is the result quoted in eq.\ (\ref{8-38}).
\bi

\no
To work out 
the angular integral (\ref{AX}),  it is more convenient to express $f (\alpha)$ (\ref{30})
in terms of its original integral representation, rather than in terms of the spherical Bessel function,
\begin{align}
f(t)=\frac{1}{2}\int\limits_{-1}^{1} dz\, (1+z^2) \cos(tz) \hk .
\end{align}
Then, the angular integral in (\ref{AX}) can be explicitly
carried out. This turns out to be also advantageous for the numerical
calculations.
\bi

\no
To carry out the angular integration in eq.\ (\ref{AX}) we use in $f (2 x \sin
\alpha)$ (\ref{30}) the
expansion
\be
\label{A4}
\cos (y \sin \alpha) & = & 
J_0 (y) + 2 \sli^\infty_{n = 1}  J_{2 n } (y) \cos 2
n \alpha  \hk ,
\ee
where $J_\nu (x)$ are the (ordinary) Bessel functions.
\bi

\no
Then we need the
following angular integrals 
\be
\label{A6}
\left. 
\il^{\frac{\pi}{2}}_0 d \alpha \cos 2 n \alpha = \frac{1}{2 n} \sin 2 n \alpha
\right|^{\frac{\pi}{2}}_0  &  = & 0 \hk \mbox{for} \hk n \neq 0 \hk , \\
\label{A7}
\il^{\frac{\pi}{2}}_0 d \alpha 2 \sin^2 \alpha \cos 2 n \alpha & = &
\il^{\frac{\pi}{2}}_0 d \alpha (2 \sin^2 \alpha - 1) \cos 2 n \alpha \nonumber\\
& = & - \il^{\frac{\pi}{2}}_0 d \alpha \cos 2 \alpha \cos 2 n \alpha = -
\frac{\pi}{4} \delta_{n, 1} \nonumber\\
 \hk .
\ee
To obtain the last relation we have inserted a zero in
form of the integral (\ref{A6}). 
With eq.\ (\ref{A6}) we find from (\ref{A4})
\be
\label{A9}
\il^{\frac{\pi}{2}}_0 d \alpha \cos (y \sin \alpha) = \frac{\pi}{2} J_0 (y) \hk
.
\ee
Analogously, we find from eq.\ (\ref{A4}) with (\ref{A7})
\be
\label{A10}
\il^{\frac{\pi}{2}}_0 d \alpha 2 \sin^2 \alpha \cos (y \sin \alpha) =
\frac{\pi}{2} \lk J_0 (y) - J_2 (y) \rk \hk .
\ee
\bi

\no
With the last two relations we
obtain for the angular integral (\ref{AX})
\be
\label{A11}
A (x) & = & \frac{1}{2} \il^1_{- 1} d z (1 + z^2) \il^{\frac{\pi}{2}}_0 d \alpha
 (1 - 2 \sin^2 \alpha) \cos (2 x z \sin \alpha) \nonumber\\
& = & \frac{1}{2} \il^1_{- 1} d z (1 + z^2) \left[ \frac{\pi}{2} J_0 (2 x z) -
\frac{\pi}{2} \lk J_0 (2 x z) - J_2 (2 x z ) \rk \right] \nonumber\\
& = & \frac{\pi}{4} \il^1_{- 1} d z (1 + z^2) J_2 (2 x z) \hk .
\ee
The remaining integral over the Bessel function can be taken using the relations
\be
\il^z_0 d t J_2 (t) & = & \il^z_0 d t J_0 (t) - 2 J_1 (z) \nonumber\\
\il^z_0 d t J_0 (t) & = & z J_0 (z) + \frac{\pi}{2} z \lk H_0 (z) J_1 (z) - H_1
(z) J_0 (z) \rk \nonumber\\
\il^1_0 d z z^2 J_2 (z y) & = & \frac{3 \pi}{2 y^2} \left[  J_2 (y) H_1 (y) -
H_2 (y) J_1 (y) \right] \hk ,
\ee
where $H_\nu (z)$ are the Struve functions.
\bi

\no
With these relations we find for the angular integral (\ref{AX}) the result
quoted in eq.\ (\ref{8-39}).
\bi

\no
We will also need the asymptotic form of $A (x)$ for small and large arguments.
This is most easily done by using the asymptotic forms of the Bessel function
$J_2 (x)$ in (\ref{A11}) (although the asymptotic form of the Struve function is
also known). One finds then the results quoted in eqs.\ (\ref{test})  and
(\ref{12-48}).
\bi

\end{appendix}

\bi

\no

\end{document}